\documentclass[reprint,nofootinbib,amsmath,amssymb,showkeys,aps]{revtex4-1}

\usepackage{graphicx}
\usepackage{dcolumn}
\usepackage{bm}
\usepackage{hyperref}
\usepackage{xcolor}

\begin{document}

\preprint{APS/123-QED}

\title{Observational Constraints on Early Coupled Quintessence}

\author{Lisa W. K. Goh$^{1}$}\email{lisa.goh@cea.fr}
\author{Joan Bachs-Esteban$^{2}$}\email{joan.bachs@tecnico.ulisboa.pt}
\author{Adri\`a G\'omez-Valent$^{3}$}\email{agvalent@roma2.infn.it}
\author{Valeria Pettorino$^{1,4}$}\email{valeria.pettorino@esa.int}
\author{Javier Rubio$^{5}$}\email{javier.rubio@ucm.es}

\affiliation{$^{1}$ Université Paris-Saclay, Université Paris Cité, CEA, CNRS, Astrophysique, Instrumentation et Modélisation Paris-Saclay, 91191 Gif-sur-Yvette, France}
\affiliation{$^{2}$Centro de Astrofísica e Gravitação - CENTRA, Departamento de Física, Instituto Superior Técnico - IST, Universidade de Lisboa - UL, Av. Rovisco Pais 1, 1049-001 Lisboa, Portugal}
\affiliation{$^{3}$INFN, Sezione di Roma 2, and Dipartimento di Fisica, Università di Roma Tor Vergata, via della Ricerca Scientifica 1, 00133 Roma, Italy}
\affiliation{$^{4}$European Space Agency/ESTEC, Keplerlaan 1, 2201 AZ Noordwijk, The Netherlands}
\affiliation{$^{5}$Departamento de Física Teórica and Instituto de Física de Partículas y del Cosmos (IPARCOS), Facultad de Ciencias Físicas, Universidad Complutense de Madrid, 28040 Madrid, Spain}

\date{\today}

\begin{abstract}
We investigate an Early Coupled Quintessence model where a light scalar mediates a  fifth force stronger than gravity among dark matter particles and leads to the growth of perturbations prior to matter-radiation equality. Using cosmological data from the $\textit{Planck}$ Cosmic Microwave Background power spectra, the Pantheon+ Type 1a Supernovae, Baryon Acoustic Oscillations, and Big Bang Nucleosynthesis, we constrain the coupling strength $\beta$ and the redshift $z_{\rm OFF}$ at which the interaction becomes effectively inactive, finding a firm degeneracy between these two parameters which holds regardless of when the scaling regime begins.
\end{abstract}

\maketitle


\section{Introduction}\label{sec:intro}
In the $\Lambda$CDM model, the growth of structure is limited to the late universe, when matter becomes the dominant cosmological component. 
This assertion fails, however, in scenarios involving additional attractive forces stronger than gravity, where structure formation can take place before matter-radiation equality \cite{Amendola_2018, Savastano_2019}.  These interactions could be mediated for instance by a sufficiently light scalar field coupled to some Beyond the Standard Model (BSM) species playing the role of dark matter, as in Coupled Quintessence (CQ) scenarios \cite{Wetterich_1994, Amendola_2000, Bettoni:2021qfs}. Interestingly enough, this type of coupling could lead to the formation of compact structures, such as primordial black holes \cite{Amendola_2018, Bonometto_2019} or primordial dark matter halos \cite{Savastano_2019}, hence affecting the cosmological structure formation history.

The effect of CQ during the Matter-Dominated Era (MDE) has been widely explored in the literature, studying linear and non-linear phenomena \cite{Amendola_2004, Bonometto_2014,  Bonometto_2017b, Bonometto_2017, Bonometto_2015, Bonometto_2017c, Sharma_2022} even with the aid of cosmological simulations \cite{Baldi_2010, Baldi_2011, Macci_2015, Casas_2016}. Additional effort has been put into constraining the model parameters with cosmological data sets \cite{Amendola_2007} while trying to alleviate cosmological tensions \cite{Barros_2019, Gomez-Valent_2020, Gomez-Valent_2022, Barros:2022bdv, Goh:2022gxo}. A particular realisation known as Growing Neutrino Quintessence has also been thoroughly examined \cite{Amendola_2008, Wetterich_2007, Ayaita_2012, Ayaita_2013, Ayaita_2016, F_hrer_2015, Mota_2008, Nunes_2011, Pettorino_2009, Wetterich_2007, Wintergerst_2010}.

In this work, we study the cosmological impact of a CQ force in the Radiation-Dominated Era (RDE). In our Early Coupled Quintessence (ECQ) scenario, a scalar field $\phi$ -- akin to dynamic dark energy (DE) -- is coupled to a BSM fermion $\psi$ -- our cold dark matter (DM) particle. As shown in \cite{Amendola_2000}, if the coupling is large enough, the system approaches an attractor solution in the RDE. Then, the scalar and fermionic energy densities evolve as the radiation one, so the density parameters $\Omega_{R}$, $\Omega_{\psi}$, and $\Omega_{\phi}$ remain constant. This scaling regime induces rapid growth of fermionic overdensities, turning swiftly non-linear \cite{Amendola_2018, Savastano_2019}.
    
In the present article, we use a modified version of the Boltzmann code \texttt{CLASS} \cite{Lesgourgues:2011re, Blas_2011} to solve the background and linear perturbation equations of the model, computing the matter and the Cosmic Microwave Background (CMB) power spectra. For the first time for ECQ, we run Monte Carlo Markov Chains (MCMC) to constrain the parameter space. 

This paper is organised as follows. In Sec. \ref{sec:FifthForce}, we review the ECQ theoretical principles. Our modified version of \texttt{CLASS} is discussed in Sec. \ref{sec:CLASS}. In Sec. \ref{sec:Data}, we describe our methodology and the datasets employed in the analyses.  In Sec. \ref{sec:Results}, we present our results and finally, in Sec. \ref{sec:Conclusions}, we draw our conclusions.  

\section{Fifth Force Interactions in ECQ}\label{sec:FifthForce}
We consider a SM extension with Lagrangian density
\begin{equation}\label{eq:Lagrangian_density}
\begin{split}
    \mathcal{L}=&\frac{M_{P}^2}{2}R+\mathcal{L}_{SM}-\frac{1}{2}\partial^{\mu}\phi\partial_{\mu}\phi-V(\phi)\\
    &+\Bar{\psi}\left(i\gamma^{\mu}\nabla_{\mu}-m_{\psi}(\phi)\right)\psi,
\end{split}
\end{equation}
where $M_{P}$ is the reduced Planck mass, $R$ the Ricci scalar, $\mathcal{L}_{SM}$ the SM Lagrangian density, $\phi$ our canonically normalised BSM scalar field, $V(\phi)$ its potential, $\psi$ our BSM fermion, and $m_{\psi}(\phi)$ its field-dependent mass through which these species interact. The interaction strength is measured by the dimensionless coupling 
\begin{equation}\label{eq:beta}
    \beta(\phi)\equiv-M_{P}\frac{\partial\text{ ln }m_{\psi}(\phi)}{\partial \phi},
\end{equation}
which causes the fifth force energy-momentum transfer
\begin{equation}\label{eq:energy-momentum_transfer}
    \nabla_{\nu}T^{\mu\nu}_{(\phi)}=\frac{\beta(\phi)}{M_{P}}T_{(\psi)}\partial^{\mu}\phi,\;\nabla_{\nu}T^{\mu\nu}_{(\psi)}=-\frac{\beta(\phi)}{M_{P}}T_{(\psi)}\partial^{\mu}\phi,
\end{equation}
where $T^{\mu\nu}_{(\phi)}$ is the energy-momentum tensor of the scalar field $\phi$, $T^{\mu\nu}_{(\psi)}$ that of the fermion $\psi$, and $T_{(\psi)}=T^{\mu\nu}_{(\psi)}g_{\mu\nu}$ its trace. 

Each choice for $\beta(\phi)$ leads to a different fermion mass behaviour and vice-versa (see Eq. \eqref{eq:beta}). A renormalisable Yukawa interaction $m_{\psi}(\phi)=m_0+g\phi$, where $m_0$ is a mass parameter and $g$ a dimensionless coupling, translates into $\beta(\phi)=-gM_p/(m_0+g\phi)$ \cite{Flores2020drq,Flores2021tmc,Domenech2021uyx,Domenech:2023afs}. Another possibility is a dilaton-like interaction $m_{\psi}(\phi)=m_0\exp{(-\beta\phi/M_p)}+m_1$ \cite{Bonometto_2017,Bonometto_2017b,Bonometto_2017c,Bonometto_2019}, with $\beta(\phi)=\beta\, m_0\exp{(-\beta\phi/M_p)}/m_\psi(\phi)$. In this case, there can be a transition from $\beta(\phi)\sim \beta$ to $\beta(\phi)\sim 0$. We aim to mimic this behaviour, also attainable through screening \cite{Savastano_2019}.

\subsection{Background evolution}\label{subsec:Background}
\begin{figure}[t!]
    \includegraphics[width =\linewidth]{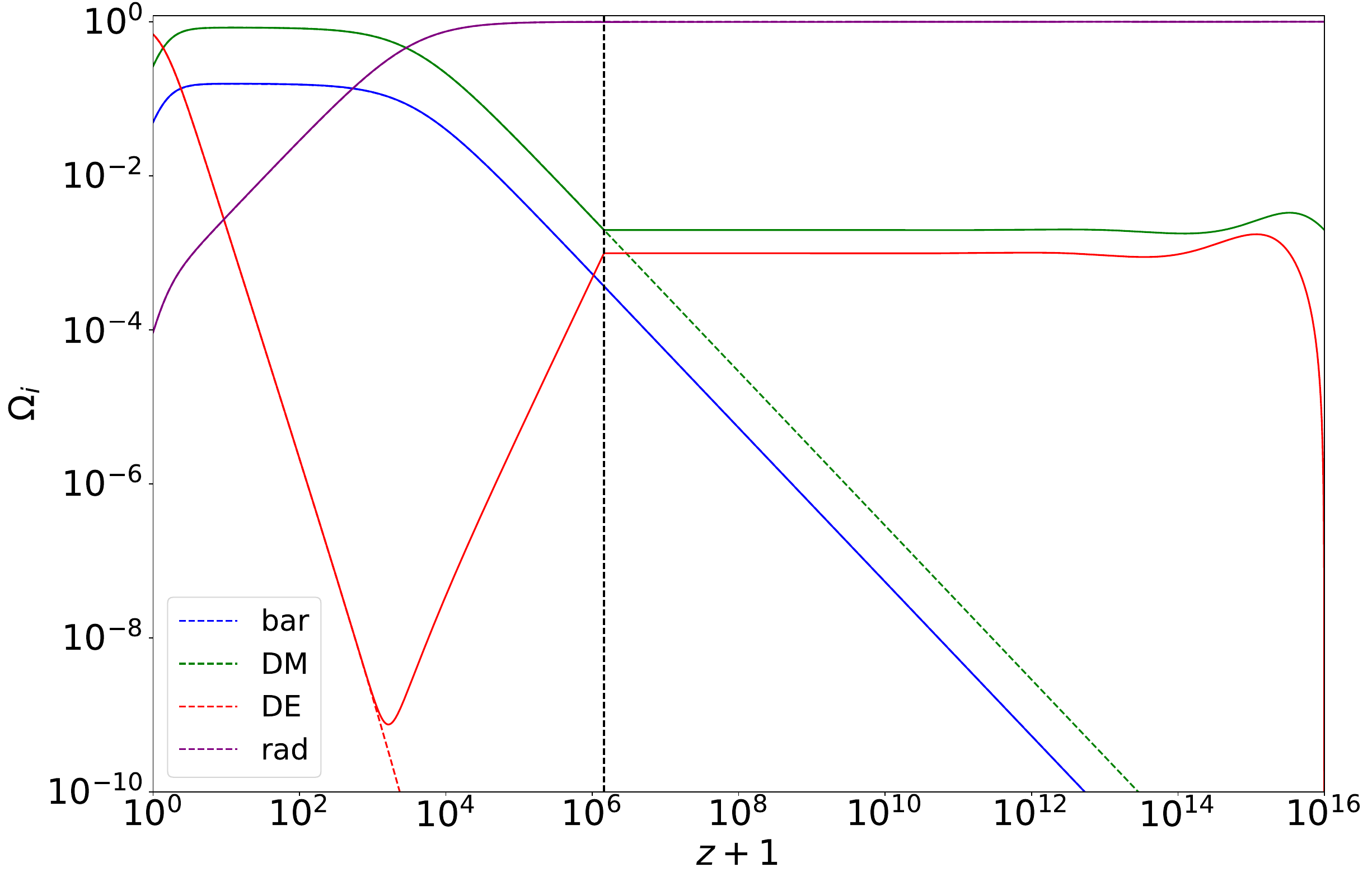}
    \caption{Background evolution of the density fractions $\Omega_i$'s of baryons (blue), DM (green), DE (red), and radiation (purple) for ECQ (solid lines) and $\Lambda$CDM (dotted lines). We fix the values of $\omega_{b}$, $V_0$, and $H_0$ to the \textit{Planck} 2018 TTTEEE+lowE+lensing $\Lambda$CDM best-fit cosmology \cite{Planck:2018vyg}, and set $\beta=13$ and $z_{\rm OFF}=10^4\beta^2=1.69\times10^6$ to retrieve the $\Lambda$CDM background evolution at $z<z_{\rm OFF}$ in the ECQ model. The grey dashed vertical line indicates $z_{\rm OFF}$. }\label{fig:best_fit_background}
\end{figure}

Let us consider a flat Friedmann-Lemaître-Robertson-Walker universe and assume that both $\psi$ and $\phi$ can be described as perfect fluids. From Eq. \eqref{eq:energy-momentum_transfer}, one can derive the following background conservation equations
\begin{align}
    \Dot{\rho}_{\phi}+3H\left(\rho_{\phi}+p_{\phi}\right)=&+\frac{\beta}{M_{P}}\left(\rho_{\psi}-3p_{\psi}\right)\Dot{\phi},\\
    \Dot{\rho}_{\psi}+3H\left(\rho_{\psi}+p_{\psi}\right)=&-\frac{\beta}{M_{P}}\left(\rho_{\psi}-3p_{\psi}\right)\Dot{\phi},
\end{align}
where the dots denote derivatives with respect to the cosmic time and $H=\dot{a}/a$ is the Hubble parameter, with $a$ the scale factor. The coupling is effectively active whenever the fermions are non-relativistic (i.e. when $\rho_{\psi}\neq3p_{\psi}$). When $\rho_\psi\gg 3p_\psi$ the system presents an attractor solution in which the scalar and fermionic fields follow the evolution of radiation for  $\beta>1/\sqrt{2}$ \cite{Amendola_2000}. In this scaling regime, the scalar field energy is dominantly kinetic and
we have \cite{Wetterich_1994, Amendola_2002, Tocchini_Valentini_2002, Bonometto_2012}
\begin{equation}\label{eq:scaling_regime_field_velocity}
    \phi'=\frac{M_P} {\beta},
\end{equation}
where the prime denotes derivatives with respect to the number of $e$-folds $N=\ln(a)$. The density parameters read (see Fig. \ref{fig:best_fit_background})
\begin{equation}\label{eq:density_parameters}
    \Omega_{\phi}=\frac{1}{6\beta^2},\quad\quad\Omega_{\psi}=\frac{1}{3\beta^2},\quad\quad \Omega_{R}=1-\frac{1}{2\beta^2}.
\end{equation}
The coupling must satisfy $\beta^2\gg 1$ to have  $\Omega_R\gg\Omega_\psi,\Omega_\phi$ in the RDE. Indeed, Big Bang Nucleosynthesis (BBN) constraints require this condition if scaling starts before BBN (see Sec. \ref{sec:Data}). 

\subsection{Perturbations evolution}
The evolution of the matter density contrast $\delta_{\psi}=\delta\rho_\psi/\rho_\psi$ can be derived from the Navier-Stokes equations \cite{Amendola_2004}. If there are no shear or rotational components in the initial velocity field, these equations condense into a single differential equation governing the non-linear growth at subhorizon scales \cite{Savastano_2019}, 
\begin{equation}\label{eq:delta_psi_growth_0}
\begin{split}
     &\delta_{\psi}''+\left(2+\frac{H'}{H}-\frac{\beta\phi'}{M_{P}}\right)\delta_{\psi}'-\frac{4}{3}\frac{\delta_{\psi}'^2}{1+\delta_{\psi}}\\
     &-\frac{3}{2}\left(Y\delta_{\psi}\Omega_{\psi}+\Omega_{R}\delta_{R}\right)\left(1+\delta_{\psi}\right)=0,
\end{split}
\end{equation}
where $Y\equiv1+2\beta^2$ is the effective coupling strength resulting from gravity plus the fifth force. Hence, for $|\beta|\approx 1$, the fifth force is of gravitational order.

In the scaling regime happening during RDE, $H'\approx-2H$. Since $\Omega_{R}\delta_{R}$ is small, if $\beta^2\gg 1$  Eq. \eqref{eq:delta_psi_growth_0} reduces to
\begin{equation}\label{eq:delta_psi_growth_1}
    \delta_{\psi}''-\delta_{\psi}'-\delta_{\psi}\left(1+\delta_{\psi}\right)-\frac{4}{3}\frac{\delta_{\psi}'^2}{1+\delta_{\psi}}=0\,,
\end{equation}
where we have used \eqref{eq:density_parameters}. The fermion overdensities are small at early times, so Eq. \eqref{eq:delta_psi_growth_1} can be linearised and analytically solved, leading to
\begin{equation}\label{eq:delta_psi_growth_2}
    \delta_{\psi}=\delta_{\psi,\text{in}}\left(\frac{a}{a_{\text{in}}}\right)^p,\,\text{with}\, p=\frac{1+\sqrt{5}}{2}\equiv\varphi\approx1.618,
\end{equation}
where the growth rate coincides with the golden ratio $\varphi$ and $a_{\text{in}}$ is the scale factor at the beginning of the scaling regime, when $\delta_\psi(a_{\rm in})=\delta_{\psi,{\rm in}}$. The scaling solution \eqref{eq:density_parameters} remains valid till the coupling is effectively switched off, either by the eventual dominance of a bare mass term or by screening effects (see paragraph preceding Sec. \ref{subsec:Background}). Afterwards, the model behaves as $\Lambda$CDM.


\section{Model Implementation}\label{sec:CLASS}
To reach the scaling solution, the energy of the scalar field must be kinetically dominated \cite{Amendola_2000}, which is achievable via a field-dependent potential with a cross-over phase or a sufficiently small constant potential. Here we mainly consider a constant potential $V=V_0\sim\mathcal{O}(M_{P}^2H_0^2)$ to produce acceleration at late times. Since this value is much lower than the critical energy density during the RDE, $\phi$ is naturally kinetically dominated and the system approaches the scaling solution \eqref{eq:density_parameters} if $\beta>1/\sqrt{2}$. 

We set the initial conditions for $\phi'$ and the DM density to the scaling solution values (recall Eqs. \eqref{eq:scaling_regime_field_velocity} and \eqref{eq:density_parameters})
\begin{equation}\label{eq:initial_conditions}
    \phi'_{\text{in}}=\frac{M_P} {\beta}, \quad\text{and} \quad\rho_{\psi,\text{in}}=\frac{\rho_{R,\text{in}}}{3\left(\beta^2-\frac{1}{2}\right)}\,,
\end{equation}
at $z_{\rm in}=10^{14}$, i.e. we assume that the system is in the scaling regime at $z_{\rm in}$. Moreover, we set $\phi_{\text{in}}=0$ since, given the considered flat potential, the dynamics of the scalar field do not depend on its initial value. 

To mimic a transition from $\beta(\phi)\sim \beta$ to $\beta(\phi)\sim 0$, we consider a simple parametrization 
\begin{equation}\label{eq:beta_model}
    \beta(z)=\frac{\beta\tanh{[s_{z}(z-z_{\rm OFF})]}+\beta}{2}\,,
\end{equation}
with $s_{z}$ a crossover rapidity that we set to a fiducial value $s_{z}=0.3$, ensuring a smooth and rapid enough transition as compared to $H^{-1}(z_{\rm OFF})$ in cosmic time, with $z_{\rm OFF}$ the redshift when the fifth force becomes negligible.

From the scaling solution \eqref{eq:density_parameters}, we expect an important degeneracy between $\beta$ and $z_{\rm OFF}$. By matching the expressions of the dark matter and radiation energy densities before and after the transition at $z_{\rm OFF}$ and considering (for now) massless neutrinos, we obtain 
\begin{equation}\label{eq:z_off_beta}
z_{\rm OFF}=-1+3\frac{\Omega^0_{\psi}}{\Omega^0_{R}}\left(\beta^2-\frac{1}{2}\right)\,\footnote{In this expression, we neglect the effect of the total neutrino mass, but we still consider relativistic neutrinos.},
\end{equation}
with the superscript $0$ denoting current quantities. A larger value for the coupling necessitates a larger redshift for it to be turned off. This is consistent with the fact that a larger $\beta$ implies a stronger fifth force (recall $Y$ from Eq. \eqref{eq:delta_psi_growth_0}) driving an intensified structure growth.
Taking values for $\Omega^0_\psi$ and $\Omega^0_R$ close to those preferred by CMB and late-time probes (see, e.g., \cite{Planck:2018vyg}), we expect $z_{\rm OFF}\sim 10^4\beta^2$, regardless of the redshift at which the scaling starts.   
We modify the Einstein-Boltzmann code \texttt{CLASS} \cite{Lesgourgues:2011re,Blas_2011} by implementing the CQ background and linear perturbation equations considering   \eqref{eq:beta_model}, with the initial conditions discussed above. Slightly different versions of this code were employed in previous analyses of CQ with $|\beta|\ll 1$ \cite{Gomez-Valent_2020, Gomez-Valent_2022,Goh:2022gxo}.

We plot the $\Lambda$CDM and ECQ density parameters $\Omega_i$'s in Fig. \ref{fig:best_fit_background}. For ECQ, we set the current energy fractions very close to the $\textit{Planck}$ fiducial $\Lambda$CDM values and fix $\beta=13$ and $z_{\rm OFF}=1.69\times10^6$, making the model lie almost exactly along the degeneracy curve \eqref{eq:z_off_beta}. We observe the presence of the scaling solution during the RDE (at $z>z_{\rm OFF}$) in the ECQ model. To better understand the individual impact of the coupling strength and the switch-off redshift, in Appendix \ref{sec:appendix_background} we plot the evolution of the density parameters and the Hubble function for a fixed $z_{\rm OFF}$ and different $\beta$ values, corresponding to ECQ models out of the degeneracy curve \eqref{eq:z_off_beta}.

\begin{figure}[t!]
    \centering
    \includegraphics[width=\linewidth]{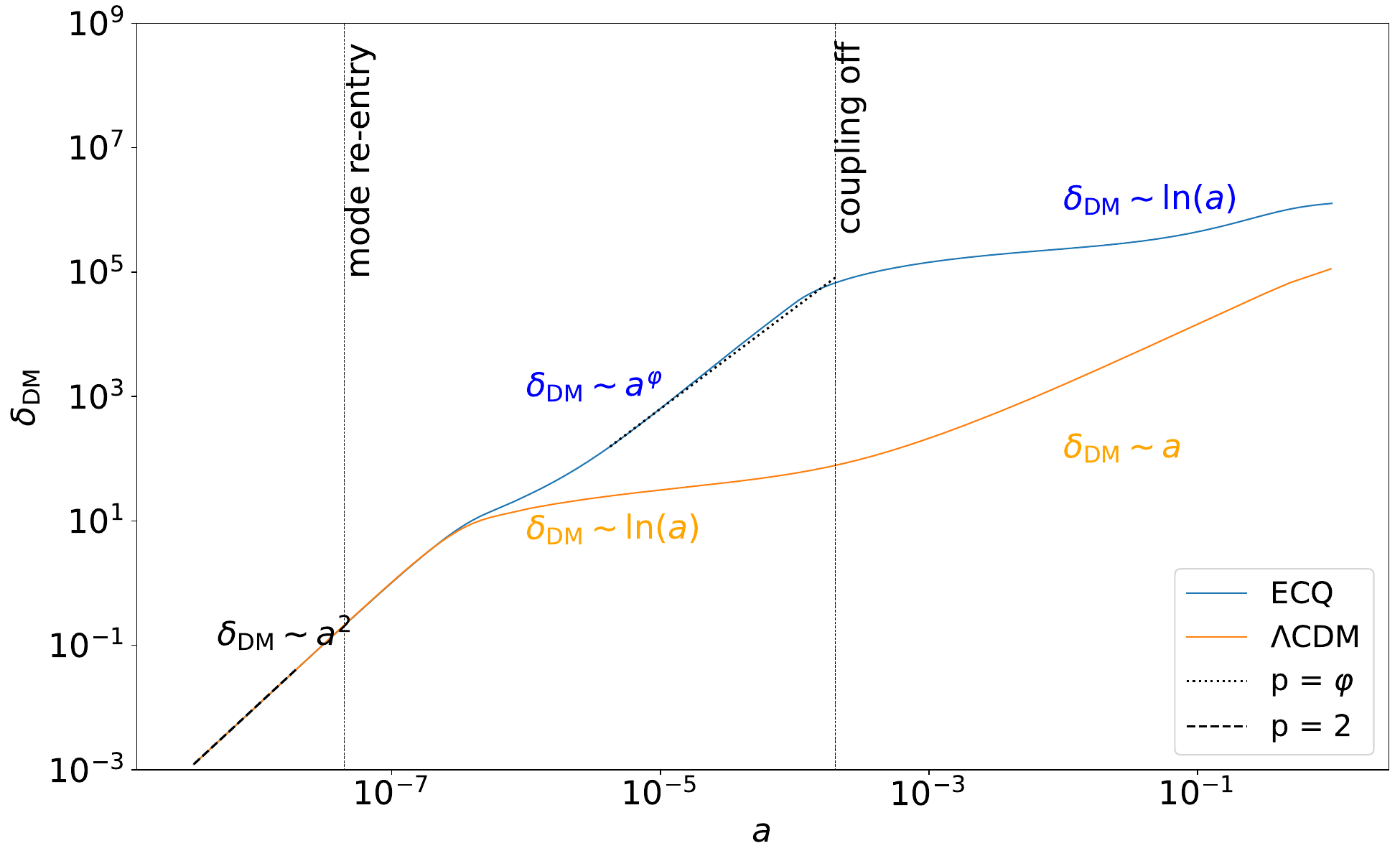}
    \caption{Evolution of DM density contrast $\delta_\mathrm{DM}$ in ECQ (blue) and $\Lambda$CDM (orange) as a function of scale factor $a$, for the scale $k = 45\,h$/Mpc. Here, we fix $n_s=0.9649$, $\ln{(10^{10}A_s)}=3.044$, $\omega_b=0.02237$, $V_0 = 2.64\cdot10^{-47}\text{ GeV}^4$, $\beta=13$, and $z_{\rm OFF}=10^4$. We include the theoretical $\delta_{DM} \sim a^2$ growth at superhorizon scales during the RDE in black dashed lines and the one during the ECQ scaling regime $\delta_{DM} \sim a^\varphi$ at subhorizon scales (Eq.  \eqref{eq:delta_psi_growth_2}) in black dotted lines. We mark with vertical lines the time of mode re-entry and when the coupling is turned off.  Notice that after turning the coupling off the universe is still strongly dominated by radiation. This is why $\delta_{\rm DM}\sim \ln(a)$ even at $a>10^{-3}$, deep inside the MDE of the  $\Lambda$CDM. This is because this particular model lies far away from the degeneracy curve \eqref{eq:z_off_beta}, such that $z_{\rm OFF}$ is too low for the ECQ model to be realistic. Thus, the recombination time would significantly change with respect to $\Lambda$CDM. The purpose of this figure is to illustrate the fulfilment of theoretical growths and the potential to enhance the power of matter fluctuations in ECQ during the RDE at sufficiently small scales.}
    \label{fig:perturbations}
\end{figure}

In Fig. \ref{fig:perturbations}, we present the growth of DM overdensities at scale $k=45 \,h$/Mpc as a function of the scale factor $a$, for both an ECQ model and $\Lambda$CDM. We notice that, for superhorizon modes in ECQ, the DM density contrast follows the $\Lambda$CDM prediction $\delta_{\rm DM}\sim a^2$. After mode re-entry, it transitions to follow the predicted golden ratio growth \eqref{eq:delta_psi_growth_2} until the coupling is turned off.

\begin{figure}[t!]
    \includegraphics[width =\linewidth]{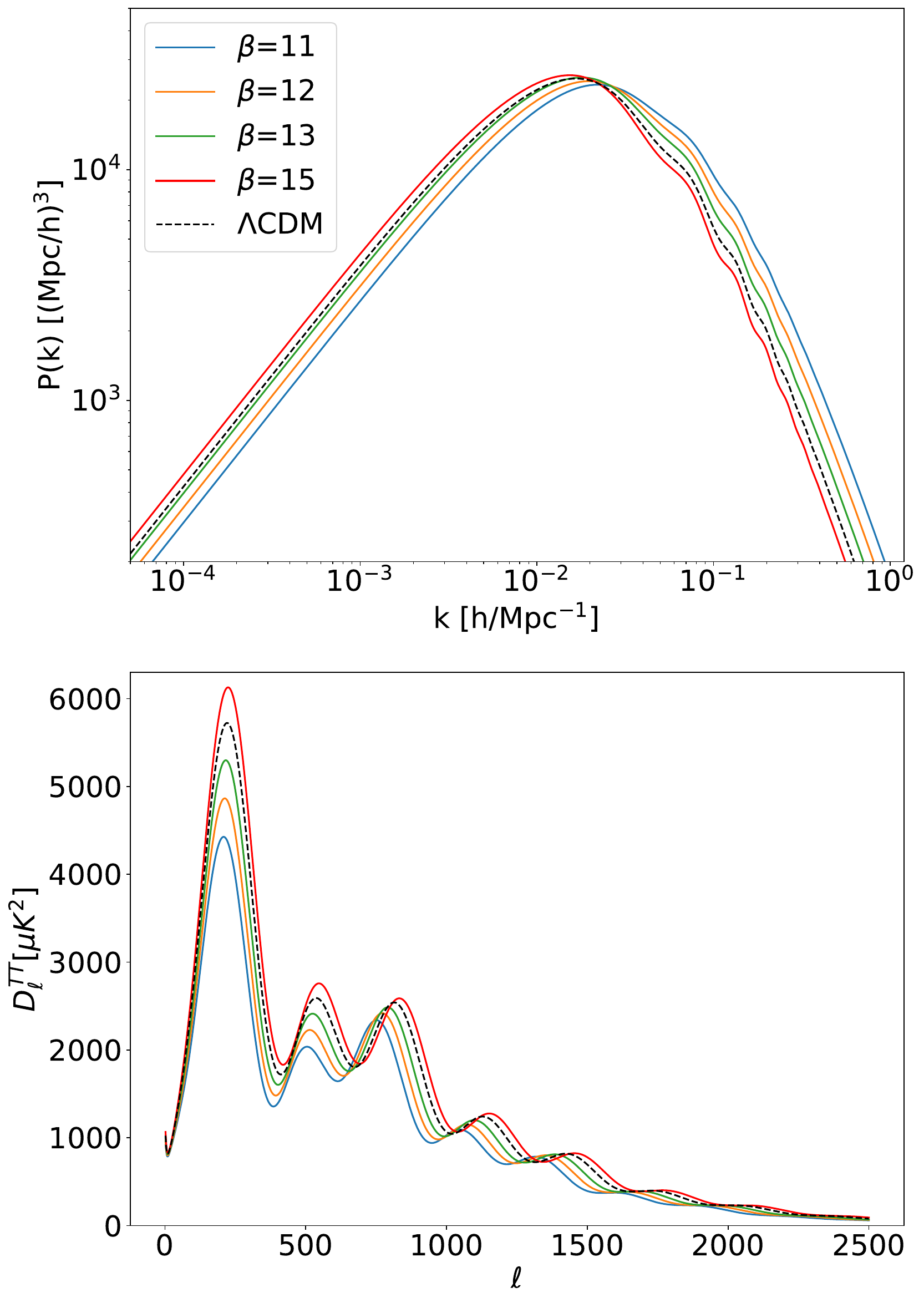}
    \caption{Matter and CMB temperature power spectra for both $\Lambda$CDM (black dashed) and ECQ (coloured solid). Here, we fix $n_s=0.9649$, $\ln{(10^{10}A_s)}=3.044$, $\omega_b=0.02237$, $\tau=0.0544$, $V_0 = 
2.64\cdot10^{-47}\text{ GeV}^4$, $z_{\rm OFF}=1.69\times10^6$ and vary $\beta=\{11,12,13,15\}$.}
    \label{fig:power_spectrum}
\end{figure}

We subsequently plot the matter and CMB temperature spectra for $\beta=\{11,12,13,15\}$ and fixed $z_{\rm OFF}=1.69{\times}10^6$ in Fig. \ref{fig:power_spectrum}. We include again the $\Lambda$CDM curves for comparison. At $z_{\rm OFF}$ we expect that only wave modes with $k\gtrsim 5\,h$/Mpc have reentered the horizon, and hence be directly affected by the fifth force. As a matter of fact, during scaling, the relation between the mode $k$ and its re-entry redshift $z$ is given by 
\begin{equation}\label{eq:mode_redshift_reentry}
    k\approx \sqrt{\frac{\Omega_R^0}{\Omega_R(\beta)}}\frac{z}{3000} \, h{\rm Mpc}^{-1}.
\end{equation}
However, smaller wave modes are also sensitive to the coupling through the changes induced at the background level. In Fig. \ref{fig:power_spectrum}, we still observe an increase of $P(k)$ at $k\gtrsim\mathcal{O}(0.1)\,h$/Mpc and a shift of its peak. This is because $\Omega_{\rm DM}$ increases and matter-radiation equality happens at larger redshifts for decreasing values of $\beta$ and a fixed $z_{\rm OFF}$. However, when $\beta$ becomes sufficiently large, $\Omega_{\rm DM}$ decreases such that $P(k)$ becomes smaller than in $\Lambda$CDM at small scales and larger at large scales (we show this for $\beta=15$).

Concerning the CMB TT spectrum, we see that the amplitude of the peaks decreases and they shift to lower multipoles for lower values of $\beta$, considering again a fixed $z_{\rm OFF}$. The decrease in amplitude can be explained by the increase in $\Omega_{\rm DM}$ at pre-recombination times. Furthermore, this also alters the baryon fraction $\rho_{\rm b}/\rho_{\rm DM}$, thereby changing the ratio between the amplitudes of the odd and even peaks. Finally, the observed shifts of the peaks to lower multipoles are again due to the increase of the DM energy density, which changes the sound horizon at the baryon-drag epoch, $r_s(z_d)$, and the angular diameter distance to the last scattering surface. The latter has a more significant relative effect, moving the peaks left.


\section{Data and Methodology}\label{sec:Data}
We conduct a Monte Carlo analysis to constrain the parameter space of the model. We sample across the suite of parameters $\{n_s,\ln{(10^{10}A_s)},\tau,\omega_{\rm b},V_0,\beta^2,z_{\rm OFF}\}$ and obtain $H_0$, $\Omega^0_{\rm DM}$, $\Omega^0_{\rm b}$, $\sigma_8$ and $S_8$ as derived parameters. We assume a flat universe with two massless neutrinos and a massive one with $m_\nu= 0.06$ eV. We sample $\beta^2$ instead of $\beta$ to expedite convergence, taking advantage of the linear relationship \eqref{eq:z_off_beta} between $z_{\rm OFF}$ and $\beta^2$. We fix the initial conditions as explained in Sec. \ref{sec:CLASS}, and assume sufficiently wide flat priors for all the input parameters. 

The relative abundance of light elements just after BBN (i.e. at $z_{\rm BBN}\sim 10^9$) places tight constraints on the expansion rate value \cite{Uzan:2010pm} and, consequently, on the density parameters $\Omega_{\psi}$, $\Omega_{\phi}$ and $\Omega_{R}$ during the scaling regime \cite{Bean:2001wt}. This translates to a rough lower bound of $\beta \gtrsim 3$ in ECQ \footnote{This restriction indicates that one cannot recover $\Lambda$CDM from ECQ by setting $\beta=0$, i.e., the two models are not nested.}. We also employ the prior upper bound $\beta<30$ since the data is insensitive to values of $\beta$ even when it is much smaller than 30. 

We also impose a flat prior on $z_{\rm OFF}$ with a range $10^5 < z_{\rm OFF} < 10^7$, where the lower bound is close to the matter-radiation equality redshift but still in the RDE and the upper bound is obtained from Eq. \eqref{eq:z_off_beta} using the prior upper bound on $\beta$. Moreover, based on BBN constraints, we consider a Gaussian prior for the baryon density, $\omega_b =0.023 \pm 0.002$, following \cite{ParticleDataGroup:2012pjm}. However, its impact on our results is minimal, as the width of its uncertainty exceeds that of the one derived from CMB data by an order of magnitude \cite{Planck:2018vyg}.

We constrain our model with the following CMB and background datasets, covering a range from low to high redshifts: 

\begin{itemize}
\item We use CMB TT, TE and EE power spectra data from the $\textit{Planck}$ 2018 data release \cite{Planck:2018vyg,Planck:2019nip}, ranging from multipoles $0 \leq \ell \leq2508$ (for TT), and $0 \leq \ell \leq1996$ (for TE and EE): more specifically,  the \texttt{simall lowl} EE and \texttt{commander lowl} TT likelihoods for  $0 < \ell < 29$, and the fast $\texttt{Plik\_lite}$ likelihood for larger multipoles, which already includes marginalization over foregrounds and residual systematics. The differences between the contours obtained with the full likelihood are marginal. 

\item We include the data on the product $H(z)r_s(z_d)$ and the ratio involving the comoving angular diameter distance $D_M(z)/r_s(z_d)$ at the three effective redshifts $z=\{0.38,0.51,0.61\}$, obtained through the reconstructed measurement of the Baryon Acoustic Oscillation (BAO) peak from the BOSS DR12 data \cite{BOSS:2016wmc}.

\item We employ the Pantheon+ dataset \cite{Scolnic:2021amr}, consisting of 1550 distinct Type Ia supernovae (SNeIa) ranging in redshift $z \in [0.001,2.26]$. We allow the absolute magnitude of the SNeIa to vary freely in the Monte Carlo analyses. This dataset gives us constraints on the current DE and DM energy fractions, which successively helps to fix the degeneracy curve in the $\beta^2-z_{\rm OFF}$ plane through Eq. \eqref{eq:z_off_beta}.

\end{itemize}

We run our version of \texttt{CLASS} to obtain the various background and perturbation quantities in the ECQ framework. Since accurate models for the non-linear power spectrum in an ECQ framework are yet to be developed, in our main analysis we opt to impose a cut in the CMB spectra at $\ell_{\rm max}=1400$ to avoid biases in the computation of the CMB lensing effects at small scales. We also set a very conservative scale cut at $k_{max}=0.1 h$/Mpc to avoid biasing the power spectrum. In Appendix \ref{sec:AdditionalRuns}, we present the results obtained in less conservative runs, considering the entire range of {\it Planck} multipoles, using \texttt{Halofit} \cite{Takahashi:2012em} without the scale cut at $k_{max}=0.1 h$/Mpc. 

We sample the posterior distributions using \textsc{cosmosis} \cite{Zuntz:2014csq}, employing a standard Metropolis-Hastings sampling algorithm \cite{1953JChPh..21.1087M,Hastings:1970aa} and stopping the sampling when the Gelman-Rubin convergence statistic $R$ fulfils $R-1<0.02$ \cite{R2:1992,R1:1997}. We analyse our chains using \texttt{GetDist} \cite{Lewis:2019xzd}. 


\section{Results}\label{sec:Results}
\begin{figure*}[t!]
\includegraphics[width=\textwidth]{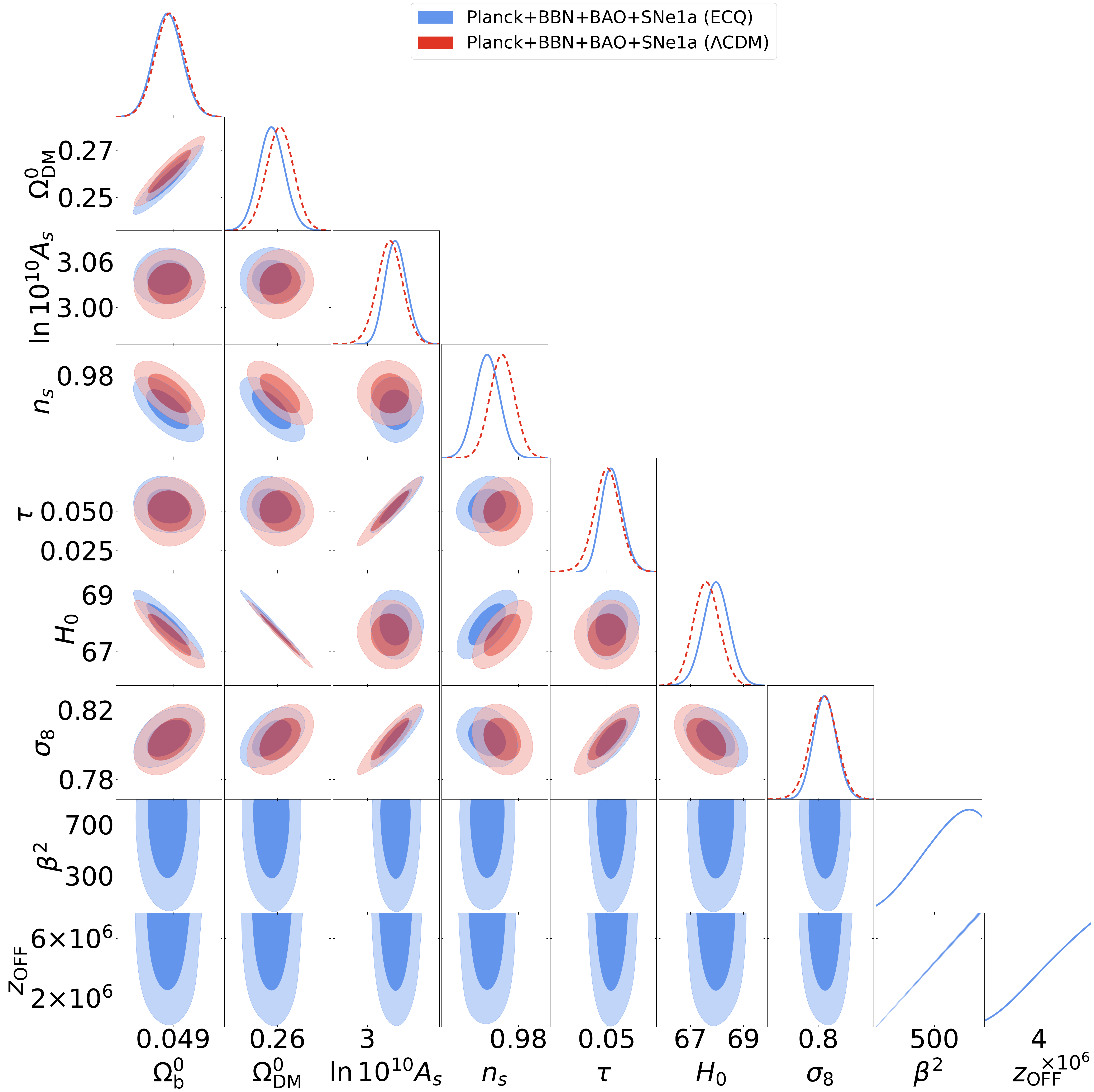}
\caption{2D contours and 1D marginalised posteriors of the main cosmological parameters for both ECQ (blue, blue solid line in 1D) and $\Lambda$CDM (red, red dashed line in 1D). The 1D posteriors for $\beta^2$ and $z_{\rm OFF}$ are essentially flat. As expected, there is a huge correlation in the $\beta^2-z_{\rm OFF}$ plane, consistent with Eq. \eqref{eq:z_off_beta}. See comments in Sec. \ref{sec:Results}.
\label{fig:corner_plot}}
\end{figure*}

We present our main results for the $\Lambda$CDM and ECQ models in Fig. \ref{fig:corner_plot}, showing the 1D and 2D marginalised posterior distributions of the most relevant parameters. We report their mean and $1\sigma$ uncertainties in Table \ref{tab:mcmc_best_fit}. We remind the reader that all the results in the main body of the manuscript have been obtained considering cuts at $\ell_{\rm max}=1400$ and at $k_{max}=0.1 h$/Mpc. We analyse the complete datasets, which include the non-linear scales, in Appendix \ref{sec:AdditionalRuns}.

\setlength{\tabcolsep}{6pt} 
\renewcommand{\arraystretch}{1.4}
\begin{table}
\centering
\begin{tabular}{lll}
\hline
\hline
\textbf{Parameter} & $\Lambda$CDM & ECQ  \\
\hline
\hline
$\Omega^0_{\rm b}$&$0.0489\pm0.0005$&$0.0488\pm0.0005$\\
$\Omega^0_{\rm DM}$&$0.261\pm0.006$&$0.257_{-0.006}^{+0.005}$\\
$\ln(10^{10}A_s)$&$3.030\pm0.016$&$3.040_{-0.015}^{+0.014}$\\
$n_s$&$0.975\pm0.004$&$0.970\pm0.004$\\
$\tau$&$0.050\pm0.008$&$0.054\pm0.007$\\
\hline
$\beta^2$&$-$&Unconstrained\\
$z_{\rm OFF}$&$-$&Unconstrained\\
$\beta^2/z_{\rm OFF}$&$-$&$(11.41_{-0.13}^{+0.11}){\times}10^{-5}$\\
\hline
$\Omega^0_\Lambda$&$0.690\pm0.006$&$0.694\pm0.006$\\
$H_0$ [km s$^{-1}$Mpc$^{-1}$]&$67.60\pm0.46$&$67.96\pm0.45$\\
$\sigma_8$&$0.803\pm0.008$&$0.804\pm0.007$\\
\hline
$\chi^2_{\rm min}$& $2000.59$ & $1998.14$\\
$\ln{B}$&$-$&$-2.29$\\
\hline
\end{tabular} 
\newline
\caption{\label{tab:mcmc_best_fit}Mean and $1\sigma$ uncertainties of $\Lambda$CDM and ECQ parameters, as well as the minimum $\chi^2$ and $\ln{B}$ values (see Sec. \ref{sec:Results}, Eq. \eqref{eq:Bayes_factor}). Since $\beta^2$ and $z_{\rm OFF}$ are individually unconstrained, we present the value of their ratio instead, which determines the slope of their degeneracy line.} 
\end{table}

\begin{figure}[ht!]
  \includegraphics[width=\linewidth]{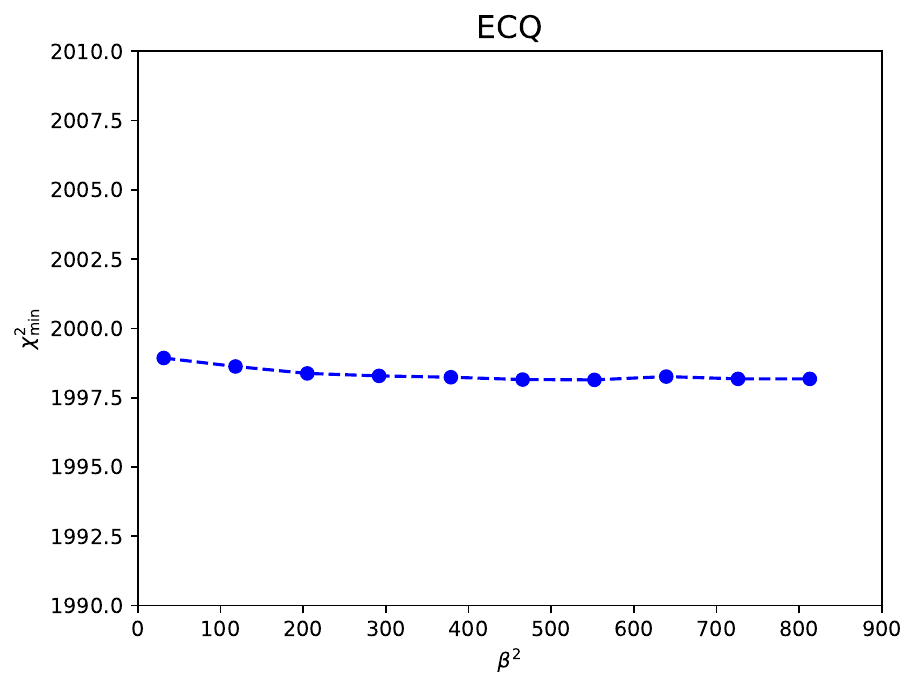}
  \caption{Minimum values of $\chi^2$ as a function of $\beta^2$, for an ECQ run considering $\textit{Planck}$ multipoles up to $\ell_{max}=1400$. We have obtained them directly from our Markov chains, applying the method of \cite{Gomez-Valent:2022hkb}. We see that the value of $\chi^2_{\rm min}$ remains roughly constant, even for those values of $\beta^2$ for which we find the peak in the 1D posterior, see Fig. \ref{fig:corner_plot}. That peak is due to volume effects.}\label{fig:beta_chi_ecq_lmax}
\end{figure}

We find the expected strong degeneracy in the $\beta^2-z_{\rm OFF}$ plane of the ECQ model, determined by Eq. \eqref{eq:z_off_beta} and the constraint on $\Omega^0_{\rm DM}$ imposed by the data. The 1D posteriors for $\beta^2$ and $z_{\rm OFF}$ are largely prior dominated. Our datasets cannot discriminate among points $(z_{\rm OFF},\beta^2)$ on the degeneracy line. There is a peak in the posterior distribution of $\beta^2$, but it is produced by volume effects, meaning that it does not lead to a significant improvement in the description of the data. This is clear from Fig. \ref{fig:beta_chi_ecq_lmax}, which shows that the minimum value of $\chi^2$ remains constant in the range of $\beta^2$ covered by the Monte Carlo.

The CMB temperature and polarisation data are only mildly sensitive to values of $\beta>3$ even if we include the entire range of {\it Planck} multipoles since in this case the fifth force affects scales that fall out of the range probed by {\it Planck}\footnote{The prior (BBN) bound of $\beta>3$, together with  Eq. \eqref{eq:z_off_beta}, lead to the lower bound $z_{\rm OFF}\gtrsim 8\times 10^{4}$, which means that even in the extreme case of $\beta=3$ only wave modes $k\gtrsim 0.25\,h{\rm Mpc}^{-1}$ are directly affected by the fifth force.}. This is clear from Fig. 19 of \cite{Planck:2018nkj}. This is also reflected in the fact that all the best-fit values of the parameters present in both models are compatible within 1$\sigma$ and that the value of $\chi^2_{\rm min}$ in ECQ is only slightly smaller than in $\Lambda$CDM (see again Table \ref{tab:mcmc_best_fit} and Fig.~\ref{fig:corner_plot}). The value of $H_0$ does not deviate considerably from the $\Lambda$CDM one, as the allowed values of $z_{\rm OFF}$ are too large to introduce significant departures from the typical $\Lambda$CDM shape of $H(z)$ at $z<10^5$ and therefore cannot induce important changes in the value of $r_s(z_d)$, which remains close to 147 Mpc. This consequently makes $H_0$ take on the $\Lambda$CDM value to keep the location of the CMB peaks stable and respect the good description of the BAO data. The model has no bearing on the cosmological tensions.

The results remain stable when we repeat the analysis including all the non-linear scales in the datasets (see Fig. \ref{fig:nonlinear-contour} in Appendix \ref{sec:AdditionalRuns}). In particular, we also get in these cases the expected degeneracy \eqref{eq:z_off_beta} in the $\beta^2-z_{\rm OFF}$ plane.

If we turn on the coupling at $z_{\rm BBN}<z<10^{14}$ with $\beta$ and $z_{\rm OFF}$ on the degeneracy line, the effects on the matter power spectrum can be softened since we decrease the interval when the fifth force is active, yet these changes affect scales that have no impact on the observables employed in this paper. Hence, in this case, we again can only constrain the slope of the degeneracy line (see Table \ref{tab:mcmc_best_fit} and contour plot in $\beta^2-z_{\rm OFF}$ plane, Fig. \ref{fig:corner_plot}). 

Lastly, we perform a model comparison by calculating the Bayes' factor of both models. This is a more robust method than simply calculating the difference in $\chi^2$ values since it additionally penalises the ECQ additional degrees of freedom. The Bayes factor is given by
\begin{equation}\label{eq:Bayes_factor}
    \ln{B}=\ln{E(M_{{\rm ECQ}}|D)}-\ln{E(M_{\Lambda{\rm CDM}}|D)}\,,
\end{equation}
with $E(M_i|D)$ the Bayesian evidence of the model $M_i$ given the dataset $D$ (see e.g. \cite{Trotta:2017wnx}). We use the code $\texttt{MCEvidence}$ \cite{Heavens:2017afc} to compute Eq. \eqref{eq:Bayes_factor} numerically. Despite that $\chi^2_{\rm min,ECQ}\lesssim \chi^2_{\rm min,\Lambda CDM}$ (see Table \ref{tab:mcmc_best_fit}), we find that $\ln{B}=-2.29$, which shows that ECQ is not preferred over the $\Lambda$CDM, according to Jeffreys' scale \cite{Kass:1995loi}. As we have explicitly verified, similar results hold also for non-constant potentials $V(\phi)$ leading to a fast-rolling field $\phi$ deep in the RDE.


\section{Conclusion}\label{sec:Conclusions}
In this paper, we have studied and constrained an ECQ model with two BSM components: a fermionic DM field $\psi$ interacting through a fifth force stronger than gravity, and a light scalar particle $\phi$ -- the interaction mediator -- which can play the role of DE.

The system admits an attractor solution whereby a scaling regime is reached during the RDE. Matter perturbations at subhorizon scales are enhanced in this period of cosmic expansion until the coupling is switched off. We have observed these features predicted by the theory in our implementation of the model into the Boltzmann code \texttt{CLASS}. We take these outcomes both as validation of our numerical Boltzmann solver and a consistency check with preceding theoretical analyses. Furthermore, we have inspected the impact of the enhanced overdensity growth in ECQ on the matter and CMB temperature power spectra and understood the physical origin of the observed changes with respect to $\Lambda$CDM. We constrain our model by employing CMB, SNeIa, and BAO datasets and running an MCMC analysis. While these are incapable of setting individual significant constraints on the model parameters, i.e. $\beta$ and $z_{\rm OFF}$, we find a distinct degeneracy in the $\beta^2-z_{\rm OFF}$ plane, fixed by the constraint on the DM parameter $\Omega^0_{\rm DM}$. We have duly constrained the degeneracy line slope.

We could improve the lower bounds on $\beta$ by using CMB data from SPT-3G \cite{SPT-3G:2021eoc} or ACT \cite{ACT:2020gnv}, which involve larger multipoles than {\it Planck}, or data from the Ly$\alpha$-forest (see e.g. Fig. 19 of \cite{Planck:2018nkj}). Nevertheless, the use of these datasets requires highly accurate modelling of the non-linear part of the matter power spectrum, something that shall be left for future work.

\begin{acknowledgments}
LG, JBE, VP, and JR acknowledge funding from Campus France and the Fundação para a Ciência e a Tecnologia (FCT) within the PESSOA bilateral cooperation program. 
LG thanks CENTRA/IST for their hospitality during her visits in October 2022 and May 2023. 
JBE acknowledges FCT for the support through Grant SFRH/BD/150989/2021 within the IDPASC-Portugal Doctoral Program. He is also grateful to CEA CNRS Universit{\'e} Paris-Saclay for the hospitality during his visits in November 2022 and December 2023.  
AGV is funded by the Istituto Nazionale di Fisica Nucleare (INFN) through the project of the InDark INFN Special Initiative: “Dark Energy and Modified Gravity Models in the Light of Low-Redshift Observations” (n. 22425/2020). 
JR is supported by a Ram\'on y Cajal contract of the Spanish Ministry of Science and Innovation with Ref.~RYC2020-028870-I.
\end{acknowledgments}

\bibliographystyle{apsrev4-2}
\bibliography{apssamp}

\appendix

\section{Non-Linear Scales MCMC Simulations}\label{sec:AdditionalRuns}

We show in this appendix the results that we find when we consider the full range of {\it Planck} multipoles in our analysis, using \texttt{Halofit} without the cut at $k_{max}=0.1 h/$Mpc. They are summarised in Fig. \ref{fig:nonlinear-contour}. 

\begin{figure}
    \centering
    \includegraphics[width=\linewidth]{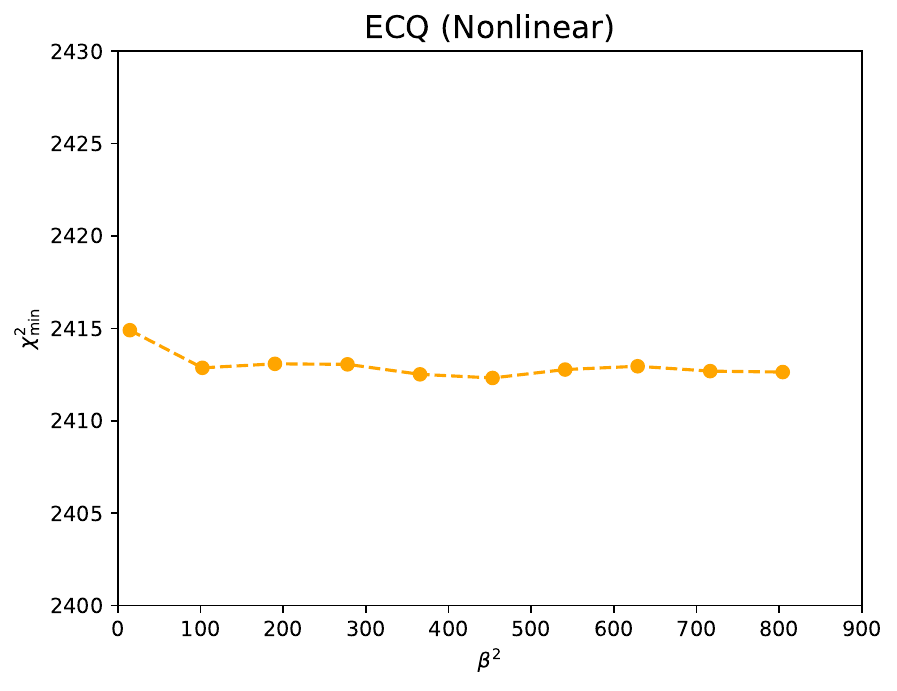}
  \caption{Minimum values of $\chi^2$ as a function of $\beta^2$, for an ECQ run considering non-linear scales ($k_{max}=1.0\,h$/Mpc and $\ell_{max}=2508$). Once again the value of $\chi^2_{\rm min}$ remains roughly constant, as in our main analysis, see Fig. \ref{fig:beta_chi_ecq_lmax}.}\label{fig:beta_chi_ecq_nonlin}
\end{figure}

\begin{figure*}[h!]
    \centering
    \includegraphics[width=\linewidth]{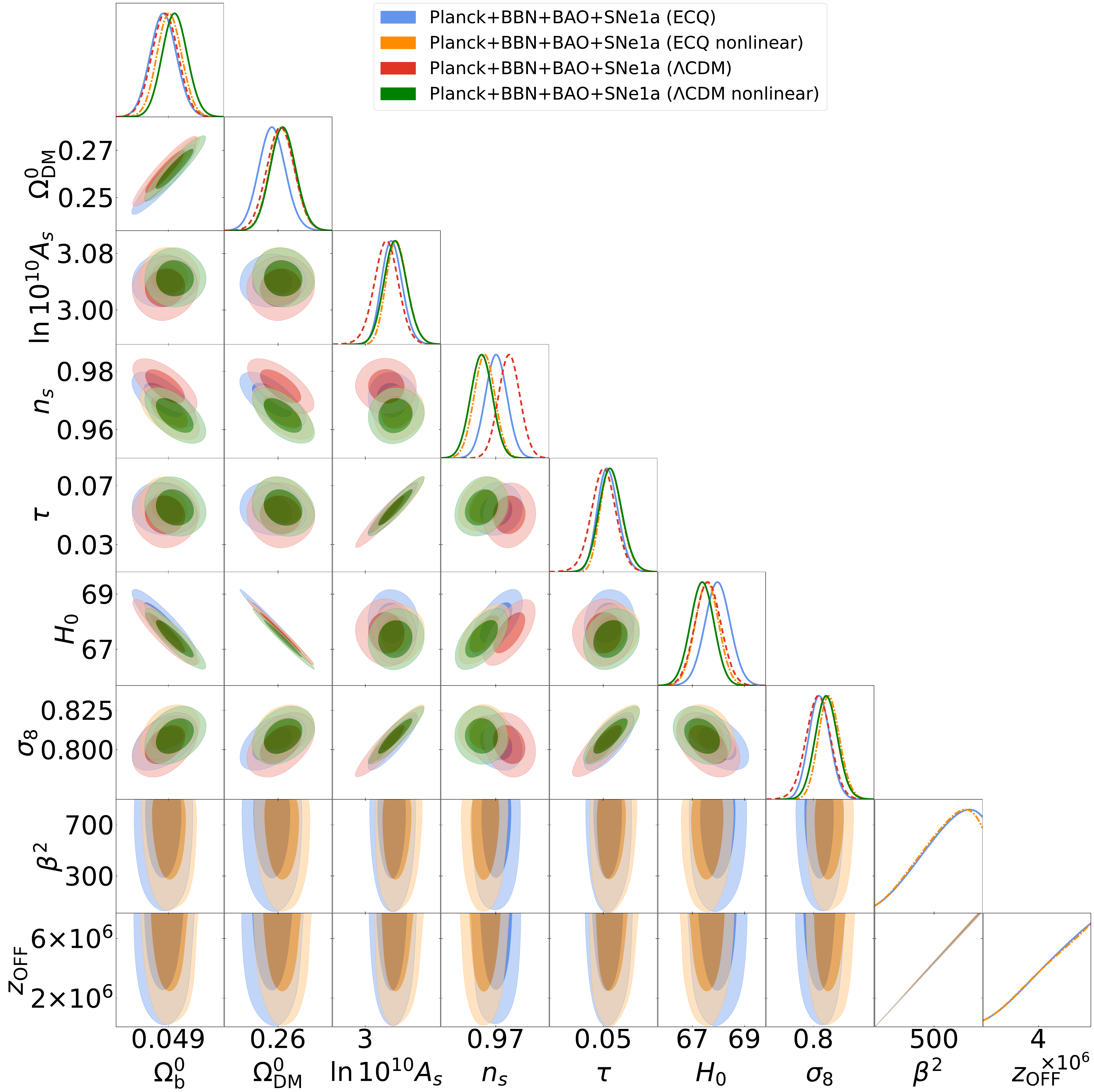}
    \caption{Comparison of the  
    2D contours and 1D marginalised posteriors of the main cosmological parameters obtained for the $\Lambda$CDM and ECQ models in the following two setups: (i) considering the linear scales, with $\ell<1400$, and using \texttt{Halofit} with the cut at $k_{max}=0.1 h/$Mpc. The results are the same as in our main analysis, displayed in Fig. \ref{fig:corner_plot}. Here we plot them in red (red dashed line in 1D) and blue (blue solid line in 1D) for the $\Lambda$CDM and ECQ, respectively; (ii) considering non-linear scales (i.e., without the cuts in $k$ and $\ell$), in green (green solid line in 1D) for $\Lambda$CDM and yellow (yellow dashed-dotted line in 1D) for ECQ. }
    \label{fig:nonlinear-contour}
\end{figure*}

As in the linear scales study (see Sec. \ref{sec:Results}), we recover the degeneracy between $\beta^2$ and $z_{\rm OFF}$ of the ECQ model. Their 1D posteriors are again prior dominated. There is again a peak in the posterior distribution of $\beta^2$, but it is produced by volume effects, meaning that it does not lead to a significant improvement in the description of the data. Once more, the minimum value of $\chi^2$ remains constant in the covered range $\beta^2$ (see Fig. \ref{fig:beta_chi_ecq_nonlin}). 

As expected, small multipoles favour larger values of $n_s$ and $H_0$, and smaller values of $\sigma_8$ (see Fig. 22 in \cite{Planck:2018vyg}). Not only do we recover this behaviour for $\Lambda$CDM but also for our ECQ model.

\section{Background Evolution}\label{sec:appendix_background}
\begin{figure*}[h!]
    \centering
    \includegraphics[width=\linewidth]{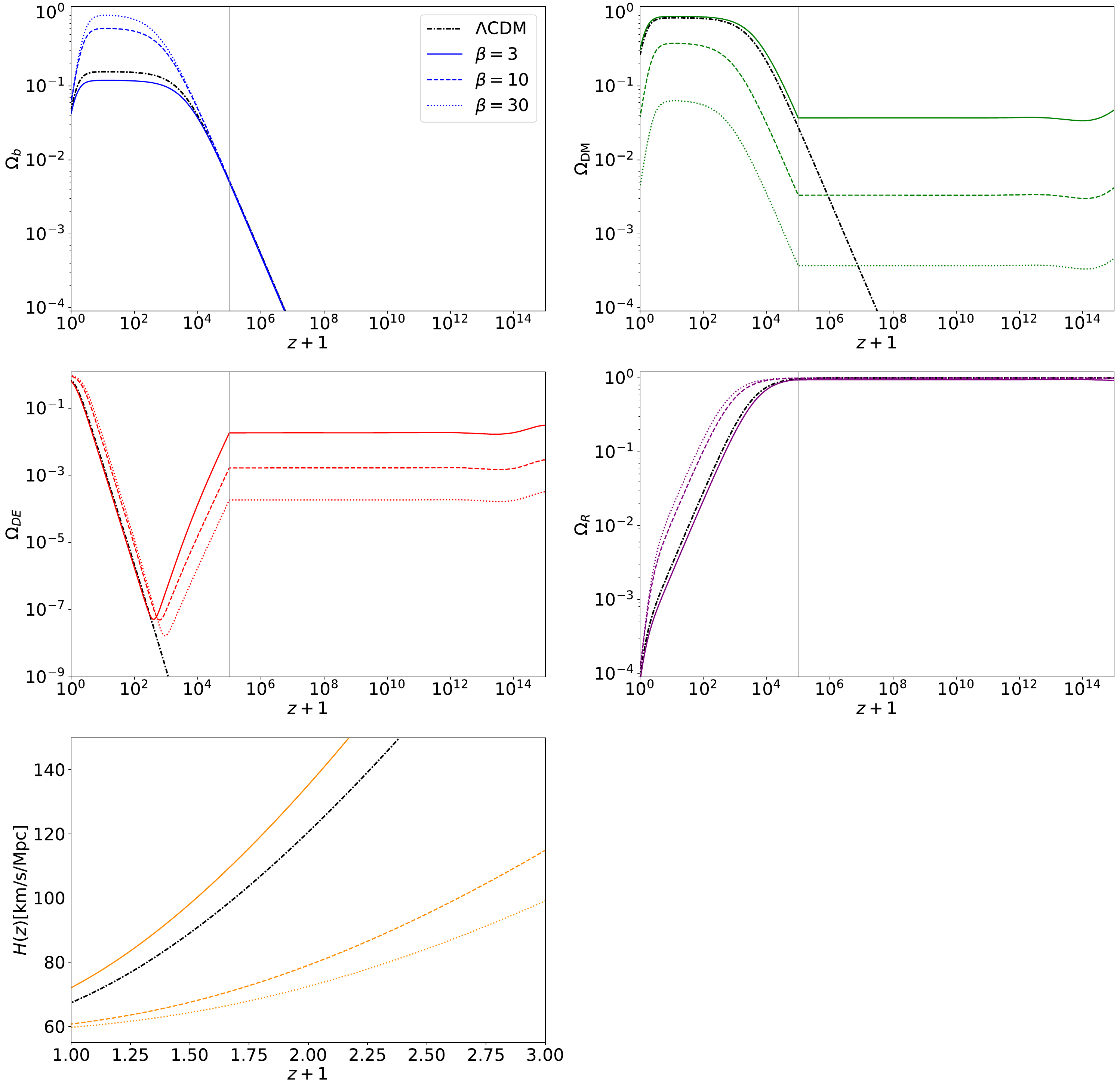}
    \caption{Evolution of the various density parameters $\Omega_i$ of (clockwise, from top left): baryons, DM, radiation and the scalar field $\phi$, for $\beta=\{3,10,30\}$. We fix $\omega_b=0.02237$, $V_0=2.64\cdot10^{-47}\text{ GeV}^4$ and $z_{\rm OFF}=10^5$. We include the $\Lambda$CDM case in black dashed-dotted lines for reference. In the last row we plot the Hubble function $H(z)$ in the redshift range $0\leq z\leq 2$.}
    \label{fig:background_appendix}
\end{figure*}

To better understand the individual effect of the coupling strength $\beta$, we present in Fig. \ref{fig:background_appendix} additional plots of the energy fractions of the various species and the Hubble function for different values of $\beta$ and fixed $z_{\rm OFF}$.

\begin{figure*}[h!]
    \centering
    \includegraphics[width=\linewidth]{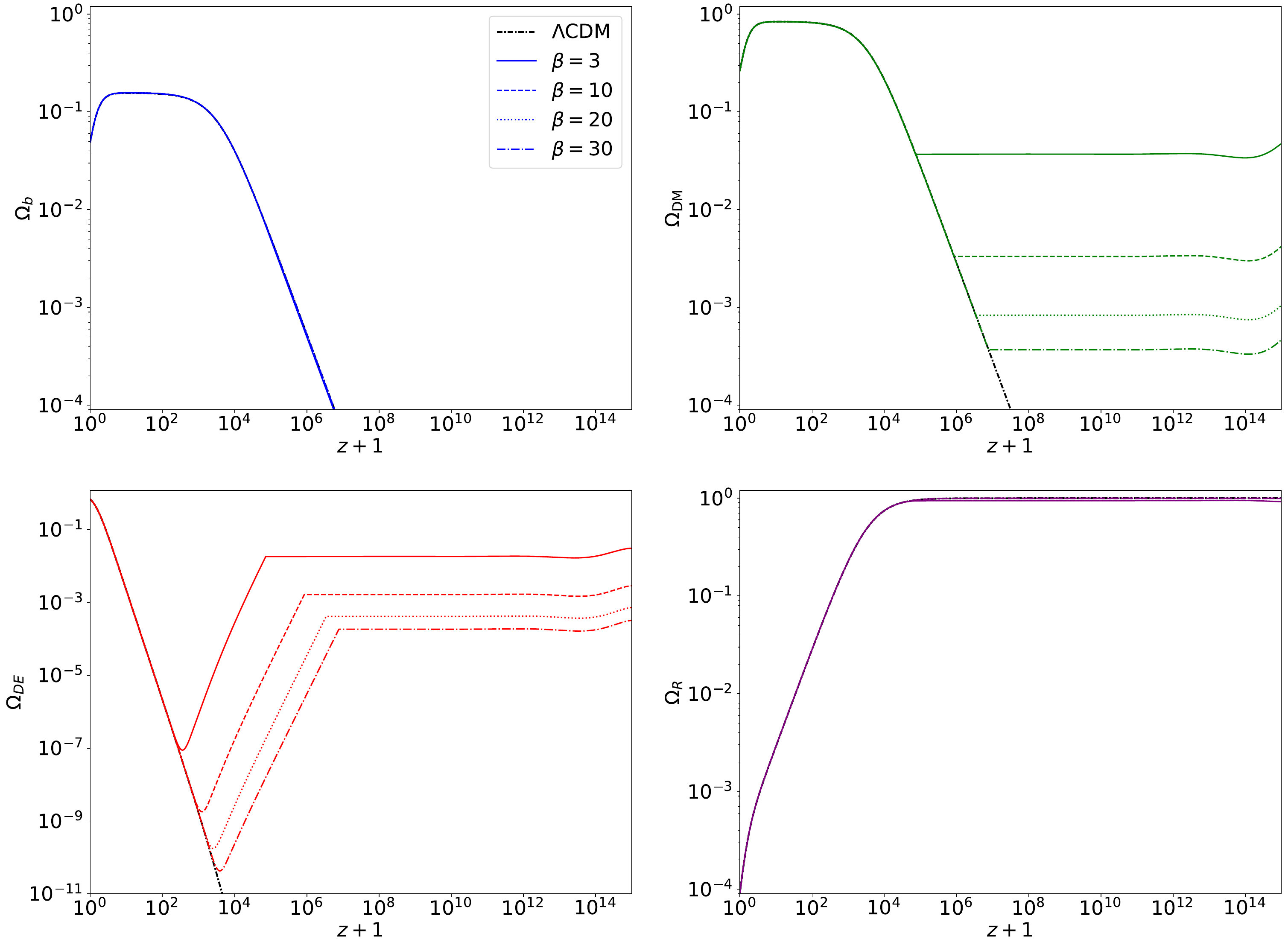}
    \caption{Evolution of the various density parameters $\Omega_i$ of (clockwise, from top left): baryons, DM, radiation and the scalar field $\phi$, for values of $\beta=\{3,10,20,30\}$ and $z_{\rm OFF}$ on the corresponding degeneracy line, Eq. \eqref{eq:z_off_beta}. We fix $\omega_b=0.02237$ and $V_0=2.64\cdot10^{-47}\text{GeV}^4$. We include the $\Lambda$CDM case in black dashed-dotted lines for reference.}
    \label{fig:background_appendix_background_degen}
\end{figure*}

\begin{figure*}[h!]
    \centering
    \includegraphics[width=\linewidth]{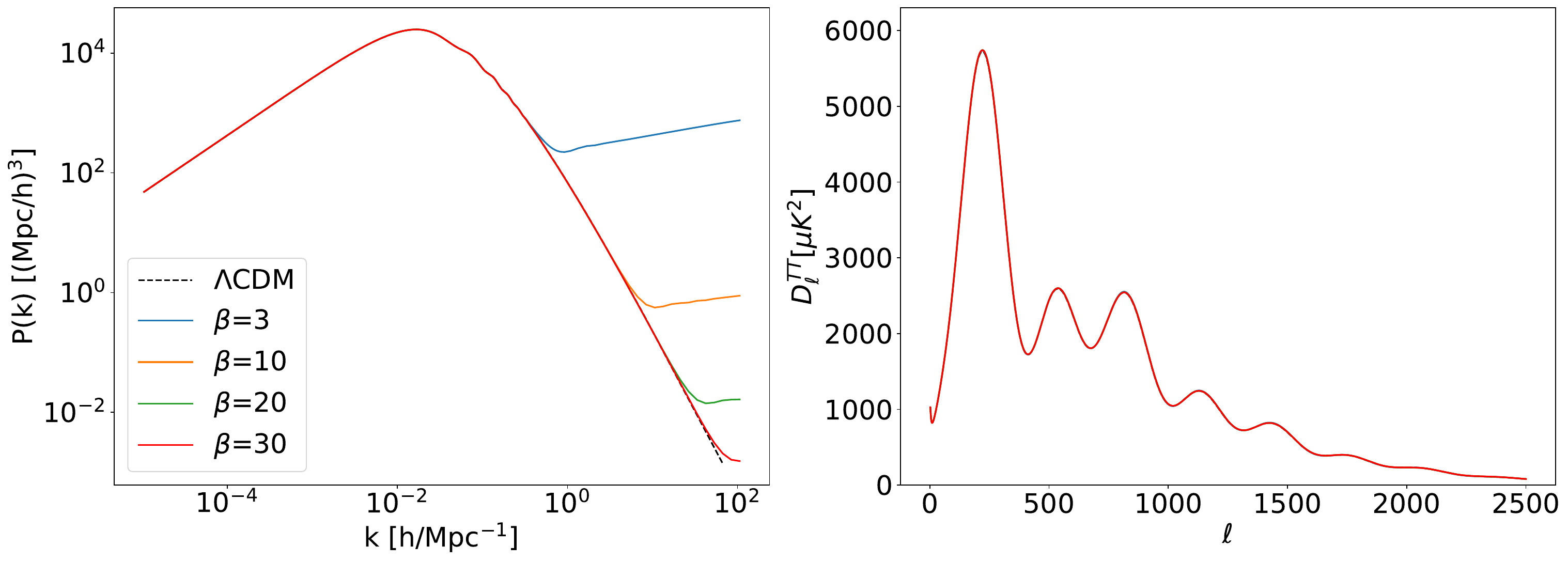}
    \caption{Matter and CMB temperature power spectra obtained using the same parameter values as in Fig. \ref{fig:background_appendix_background_degen}, i.e. using again values of $\beta$ and $z_{\rm OFF}$ lying on the degeneracy line. We include the $\Lambda$CDM case in black dashed-dotted lines for reference. In the left plot, we show a larger range of $k$'s than in Fig. \ref{fig:power_spectrum} to appreciate the impact of the large values of $\beta\gtrsim \mathcal{O}(10)$ on $P(k)$.}
    \label{fig:background_appendix_pk_degen}
\end{figure*}

In Figs. \ref{fig:background_appendix_background_degen} and \ref{fig:background_appendix_pk_degen} we plot the background evolution of the $\Omega_i$'s and the matter and CMB temperature power spectra, respectively, for values of $\beta$ and $z_{\rm OFF}$ along the degeneracy line, Eq. \eqref{eq:z_off_beta}. In Fig. \ref{fig:background_appendix_pk_degen} we see how the matter power spectrum diverges at higher $k$'s for increasing values of $\beta$. However, the datasets employed in this work (cf. Sec. \ref{sec:Data}), including the CMB spectra, are not sensitive to these small-scale effects. This is clear from the right plot of Fig. \ref{fig:background_appendix_pk_degen}. 

\end{document}